\documentclass[fleqn,twoside]{article}
\usepackage{espcrc2}

\usepackage{graphicx}
\usepackage[figuresright]{rotating}

\usepackage{epsfig,xspace}

\newcommand{\AmS}{{\protect\the\textfont2
  A\kern-.1667em\lower.5ex\hbox{M}\kern-.125emS}}

\title{Triviality and the Higgs mass lower bound}

\author{K.~Holland\address[UCSD]{Dept.~of Physics,
        University of California San Diego, 9500 Gilman Drive, 
        La Jolla CA 92093-0319, USA}%
        \thanks{Research supported by the DOE under 
	        grant DOE-FG03-97ER40546.}
	}
       
\begin{document}

\begin{abstract}
In the minimal Standard Model, it is commonly believed that the Higgs 
mass cannot be too small, otherwise Top quark dynamics makes the Higgs potential 
unstable. Although this Higgs mass lower bound is relevant for current 
phenomenology, we show that the Higgs vacuum instability in fact does not exist 
and only appears when treating incorrectly the cut-off in the renormalization of a 
trivial theory. We also demonstrate how to calculate correctly the regulator-dependent 
Higgs mass lower bound.

\vspace{1pc}
\end{abstract}

\maketitle

\section{VACUUM INSTABILITY}

\begin{figure}[htb]
\vspace{9pt}
\epsfig{figure=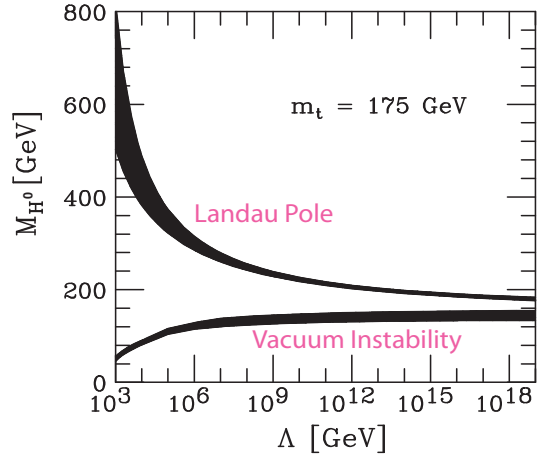,height=7cm,width=6cm,angle=90}
\caption{Upper and lower SM $m_{\rm Higgs}$ bounds \cite{Hagiwara:2002fs}.}
\label{fig:PDG}
\end{figure}

In this talk I report on some recent work which was done in
collaboration with Julius Kuti \cite{Holland:2004}.
The as-yet unobserved Higgs boson plays a crucial role in determining the
threshold of new physics beyond the Standard Model (SM). Precision Electroweak 
measurements indicate that, if the SM is correct, the Higgs is light with a mass 
$m_{\rm Higgs}=114^{+69}_{-45}$~GeV \cite{LEP04}. Even if one excludes 
these experimental constraints, it is widely believed on theoretical grounds that 
the Higgs mass must lie within a certain range for the SM to be an acceptable field 
theory. Fig.~\ref{fig:PDG} shows the current phenomenological upper and lower 
$m_{\rm Higgs}$ bounds as a function of $\Lambda$, the energy scale at which physics 
beyond the SM must set in \cite{Hagiwara:2002fs}. Upper and lower bounds are very 
important for two reasons. Firstly, they tell us where we should look for the Higgs. 
For example, according to Fig.~\ref{fig:PDG}, the SM cannot sustain a Higgs as heavy 
as 1~TeV. Secondly and more importantly, if the Higgs is observed, knowing its mass would 
tell us the maximum energy scale up to which the SM can be valid. According to this plot, 
the lower bound is the relevant one for current phenomenology. For example, taking the 
preferred value $m_{\rm Higgs}=114$~GeV, the SM is at most valid up to around 
$100-1000$~TeV. 
The phenomenological lower bound is based on the instability of the 
Higgs potential if $m_{\rm Higgs}$ is too small. We will show that this vacuum instability,
like the Landau pole, is fake. Remarkably, just like the upper bound, a new
regulator-dependent lower bound emerges from the triviality of the theory.
An earlier version of this work was presented in \cite{Holland:2003jr}.

The apparent vacuum instability can be seen in a Higgs-Yukawa model of a single 
real scalar field (Higgs) coupled to $N_{\rm F}$ degenerate fermions (Top quarks). 
The 1-loop Higgs effective potential $U_{\rm eff}$ is 
calculated by summing an infinite series of diagrams, giving 
\begin{eqnarray}
U_{\rm eff} &=& V + 1/2 \int_k \ln[k^2 + V''] \nonumber \\
&-& 2 N_{\rm F} \int_k \ln[k^2 + y^2\Phi^2],
\nonumber \\ 
V &=& m^2 \Phi^2/2 + \lambda \Phi^4/24. 
\label{eq:Ubare}
\end{eqnarray}
The fermion contribution is negative, 
due to the minus sign associated with every fermion loop. Regulating the integrals
and adding counterterms to absorb the divergences in the normal fashion, the 
renormalized effective potential is
\begin{eqnarray}
U_{\rm eff} &=& V + \{(V'')^2/64 \pi^2\}\{ \ln[V''/\mu^2] - 3/2 \}
\nonumber \\
&-& \hspace{-0.3cm}\{N_{\rm F} y^4 \Phi^4/16 \pi^2\} \{ \ln[y^2 \Phi^2/\mu^2] - 3/2 \}.
\label{eq:Ucontinuum}
\end{eqnarray}
For $\Phi$ large, the negative fermion contribution dominates if
$\lambda^2 < 16 N_{\rm F} y^4$ and the potential appears unstable, it
no longer has its ground state at $\Phi=v$. For fixed Yukawa coupling
$y$ (i.e.~$m_{\rm Top}/v$), if we require that $U_{\rm eff}$ be stable,
this gives a lower bound for $\lambda$ and hence $m_{\rm Higgs}$
(at tree-level the relation is 
$\lambda=3m_{\rm Higgs}^2/v^2$) \cite{Krive:1976sg}.

\begin{figure}[thb]
\vspace{9pt}
\epsfig{figure=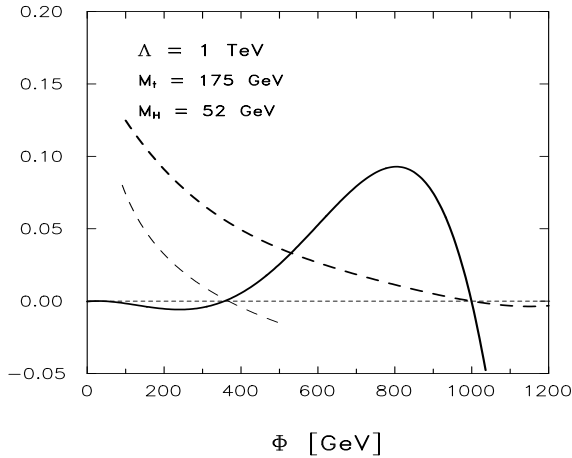,height=6cm,width=7.5cm,angle=0}
\caption{Higgs potential (solid line) and $\lambda(\mu=\alpha \Phi)$ 
        (thick dashed line) for $m_{\rm Higgs}=52$~GeV \cite{Casas:1996aq}.}
\label{fig:Espinosa}
\end{figure}

To eliminate the large logs in Eq.~\ref{eq:Ucontinuum}, the RG-improved
Higgs effective potential in the Standard Model has been calculated to
two loops \cite{Casas:1996aq,Altarelli:1994rb}. Fig.~\ref{fig:Espinosa} shows this 
effective potential for a particular choice of $m_{\rm Top}$ and $m_{\rm Higgs}$. 
If Fig.~\ref{fig:Espinosa} were representing the true behavior in the SM
for $m_{\rm Higgs} = 52$~GeV, the ground state at $\Phi=v= 246$~GeV
could be preserved only by changes in the shape of $U_{\rm eff}$ from new
physics beyond the SM around $\Phi \sim 1$~TeV. If $m_{\rm Higgs}$ is increased,
the scale of the required new physics is also increased as shown in 
Fig.~\ref{fig:PDG}.
Fig.~\ref{fig:Espinosa} also shows the running coupling 
$\lambda(\mu=\alpha \Phi)$. For a particular choice of $\alpha$, the
instability of the potential coincides with $\lambda(\mu = \alpha\Lambda)=0$. 
The lower bound in Fig.~\ref{fig:PDG} is the smallest $m_{\rm Higgs}$ value 
such that the instability appears at $\Phi=\Lambda$. The finite thickness
of the lower bound is an estimate of the uncertainty of the theoretical 
calculation. In a renormalizable field theory, one expects that this 
uncertainty can, in principle, be reduced.

\begin{figure}[htb]
\vspace{9pt}
\epsfig{figure=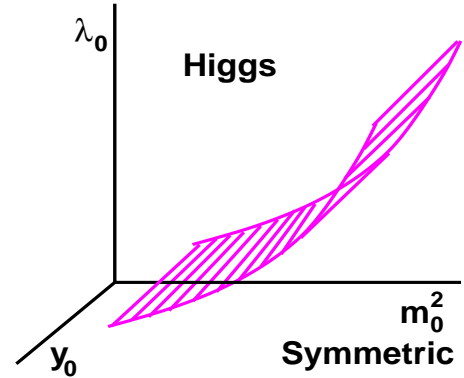,height=5cm,width=6cm,angle=0}
\caption{Phase diagram of Higgs-Yukawa model.}
\label{fig:phase}
\end{figure}  

\section{VACUUM IS STABLE}
\begin{figure}[thb]
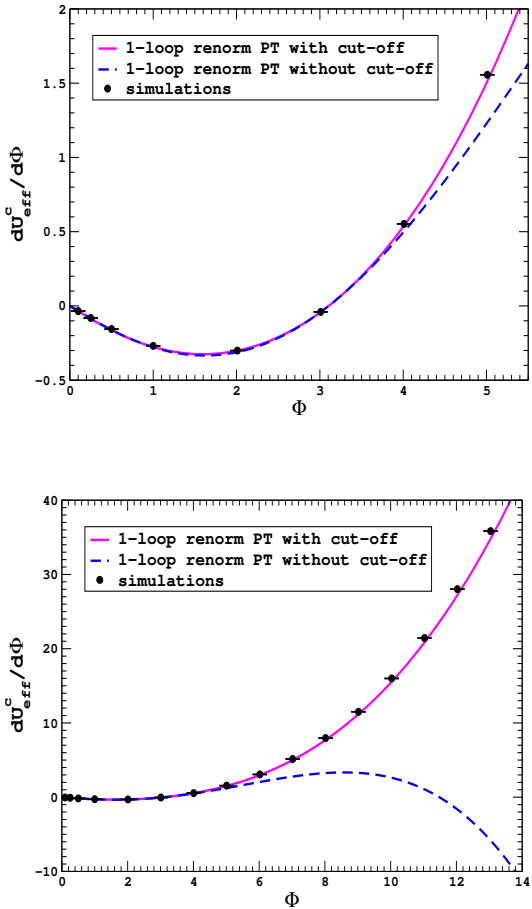

\vspace{9pt}
\epsfig{figure=dUdphi_simul_pert_3.eps,height=5.5cm,width=7cm,angle=0}
\phantom{.}\vspace{1cm}
\epsfig{figure=dUdphi_simul_pert_4.eps,height=5.5cm,width=7cm,angle=0}
\caption{$dU^c_{\rm eff}/d\Phi$ shown on two different scales.}
\label{fig:dUdphi1}
\end{figure}
The effective potential $U_{\rm eff}$ can be calculated non-perturbatively
via lattice simulations as shown by Kuti and Shen \cite{Kuti:1987bs}. We 
will concentrate on the Higgs-Yukawa model of one real scalar field coupled to 
$N_{\rm F}$ degenerate fermions. Fig.~\ref{fig:phase} shows the phase diagram 
of the lattice-regulated theory as a function of the bare couplings 
$\lambda_0, y_0$ and $m^2_0$. The lattice spacing is $a$ and the cut-off of 
the theory is $\Lambda=\pi/a$, the maximum-allowed momentum. The Higgs and 
symmetric phases are separated by a critical surface where the vacuum expectation 
value and physical masses vanish, $ va, m_{\rm ph}a 
\rightarrow 0$~(all dimensionful quantities are calculated in lattice-spacing units). 
Close to the critical surface on the Higgs side, $\Lambda/m_{\rm ph}=\pi/m_{\rm ph}a$ 
and $\xi/a=1/m_{\rm ph}a$ are large ($\xi$ is the correlation length) and the 
theory is very close to the continuum limit where the cut-off is sent to infinity. 
In this scaling region, effects due to the finite cut-off are expected to be small.

For a given action $S[\phi]$, the constraint effective potential $U^c_{\rm eff}$ 
in a finite volume $\Omega$ is
\begin{eqnarray}
\exp(-\Omega U^c_{\rm eff}(\Phi)) &=& \int [D\phi]
\delta(\Phi - 1/\Omega \sum_x \phi(x)) \nonumber \\
&& \phantom{.} \hspace{1cm}\cdot\exp(-S[\phi]),
\label{eq:Uconstraint}
\end{eqnarray}
where the delta-function constrains the average of $\phi$ to a fixed 
value $\Phi$ . The constraint effective potential $U^c_{\rm eff}$ has 
an absolute minimum at non-zero $\Phi$ in the Higgs phase even at finite 
volume $\Omega$, as the constraint potential is not convex 
\cite{O'Raifeartaigh:1986hi}. This allows the Higgs and symmetric
phases to be clearly distinguished at finite volume.

We examine the Higgs-Yukawa model with 2 flavors of staggered lattice fermions 
(this corresponds to $N_{\rm F}=8$ fermions in the continuum limit). The
derivative of the constraint effective potential is
\begin{eqnarray}
dU^c_{\rm eff}/d\Phi &=& m^2_0 \Phi + 1/6 \hspace{0.1cm}\lambda_0 \langle \phi^3 \rangle_\Phi
\nonumber \\
&-& 2y_0 \langle {\rm Tr}(D^{-1}[\phi]) \rangle_\Phi,
\label{eq:dUdphi}
\end{eqnarray}
where $D$ is the Dirac operator and $\langle ... \rangle_\Phi$ denotes
expectation values where $\Phi = 1/\Omega \sum_x \phi(x)$ is held
fixed. In Fig.~\ref{fig:dUdphi1} 
we plot $dU^c_{\rm eff}/d\Phi$ for a particular choice of bare couplings 
in the Higgs phase. From the lattice simulations, we see that 
$dU^c_{\rm eff}/d\Phi$ vanishes at 
$\Phi =0$~(a local maximum) and $\Phi \approx 3.1$~(the absolute minimum). 
There is no indication that the potential becomes unstable at large $\Phi$. 

We also compare the results of the simulations with 1-loop renormalized 
perturbation theory. Adding explicitly the counterterms and keeping the
cut-off $\Lambda$ finite, the 1-loop constraint effective potential is
\begin{eqnarray}
U^c_{\rm eff} &=& V + 1/2 \int_{k \ne 0}^{\Lambda} \ln[1 + V''/k^2]
\nonumber \\
&-& 2 N_{\rm F} \int_{k \ne 0}^{\Lambda} \ln[1 + y^2 \Phi^2/k^2]
\\
&-& 1/2 \int_{k \ne 0}^{\Lambda} \{ V''/k^2 - (V'')^2/2[k^2 + \mu^2]^2 \}
\nonumber \\
&+& 2 N_{\rm F} \int_{k \ne 0}^{\Lambda} \{ y^2 \Phi^2/k^2
- y^4 \Phi^4/2[k^2 + \mu^2]^2 \}. \nonumber
\label{eq:Urenorm}
\end{eqnarray}
Naively taking the cut-off $\Lambda \rightarrow \infty$ gives the continuum
effective potential as in Eq.~\ref{eq:Ucontinuum}. In Fig.~\ref{fig:dUdphi1} 
we see that for small $\Phi$, continuum renormalized perturbation theory 
agrees well with the simulation results, even though the cut-off is naively 
sent to infinity. However for larger $\Phi$, continuum perturbation
theory incorrectly predicts that the ground state is unstable. In contrast, 
renormalized perturbation theory {\em with a finite cut-off} is in perfect 
agreement with the non-perturbative simulation results for all values of 
$\Phi$. The ground state of the theory is in fact stable.

\section{TRIVIALITY}

The Higgs potential only appears unstable when the cut-off is incorrectly
sent to infinity. The standard renormalization procedure of adding
counterterms and removing the cut-off fails in a trivial theory. A
quantum field theory is defined by a set of bare couplings and a regulator.
A theory is trivial if the renormalized couplings vanish when the regulator 
is removed, for any choice of bare couplings. In this situation, 
the cut-off must remain finite to have a non-trivial interacting theory.

\begin{figure}[thb]
\vspace{9pt}
\epsfig{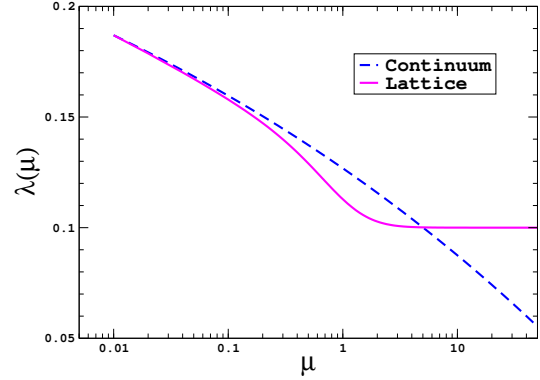}
\phantom{.}\vspace{1cm}
\epsfig{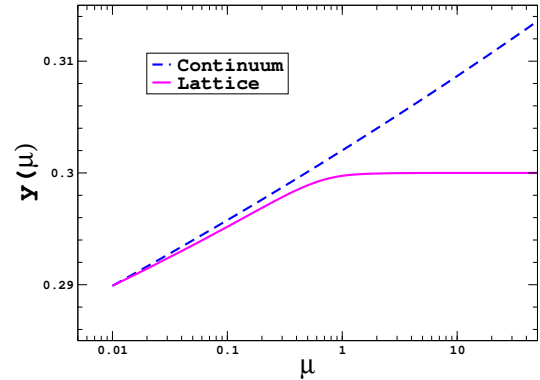}
\caption{Renormalized couplings $\lambda(\mu)$ and $y(\mu)$.}
\label{fig:RG}
\end{figure}

We will show how standard renormalization fails in the trivial Higgs-Yukawa 
model in the large-$N_{\rm F}$ limit. In terms of the bare fields and couplings, 
the Lagrangian is
\begin{eqnarray}
{\cal L} &=& m^2_0 \phi^2_0/2 + \lambda_0 \phi^4_0/24 + y_0 \phi_0
\bar{\Psi}^a_0 \Psi^a_0 + {\rm K.E.} \nonumber \\
&=&  m^2_0 Z_{\phi} \phi^2/2 + \lambda_0 Z_{\phi}^2 \phi^4/24 \nonumber \\
&+& y_0 Z_{\phi}^{1/2} Z_{\Psi} \phi \bar{\Psi}^a \Psi^a + {\rm K.E.} 
\nonumber \\
&=& (m^2 + \delta m^2) \phi^2/2 + (\lambda + \delta \lambda) \phi^4/24 
\nonumber \\
&+& (y + \delta y) \phi \bar{\Psi}^a \Psi^a + {\rm K.E.}
\label{eq:bareL}
\end{eqnarray}
where $a=1,...,N_{\rm F}, \phi_0 = Z_{\phi}^{1/2} \phi, 
\Psi_0 = Z_{\Psi}^{1/2} \Psi$ and $\delta m^2, \delta \lambda$ and $\delta y$ 
are the counterterms. In the limit $N_{\rm F} \rightarrow \infty$, 
only Feynman diagrams with fermion loops contribute, hence $Z_{\Psi}=1$. The
bare and renormalized couplings are related as
\begin{eqnarray}
&& m_0^2 Z_{\phi} = m^2 + \delta m^2, \hspace{1cm} 
\lambda_0 Z_{\phi}^2 = \lambda + \delta \lambda, \nonumber \\
&& y_0 Z_{\phi}^{1/2} = y + \delta y, 
\label{eq:barerenorm}
\end{eqnarray}
where the counterterms are
\begin{eqnarray}
\delta m^2 &=& 4 N_{\rm F} y^2 \int_k^\Lambda 1/k^2, \nonumber \\
\delta y &=& 0, \nonumber \\
\delta \lambda &=& -24 N_{\rm F} y^4 \int_k^\Lambda 1/[k^2 + \mu^2]^2,
\label{eq:counter}
\end{eqnarray}
and the wave-function renormalization is
\begin{eqnarray}
G^{-1}_{\phi}(p^2) &=& p^2 + m^2_{\rm Higgs} - \Sigma(p^2), \nonumber \\
Z_{\phi}^{-1} &=& dG^{-1}_{\phi}/dp^2|_{p^2=0}, \nonumber \\
\Sigma(p^2) &=& - 4 N_{\rm F} y^2 \int_k^\Lambda 
[\mu^2 - k\cdot(k-p)]/ \nonumber \\
&& \phantom{.} \hspace{0.5cm} \{ [k^2 + \mu^2][(k-p)^2 + \mu^2] \}. 
\label{eq:zphi}
\end{eqnarray}
Note that we keep the regulator cut-off $\Lambda$ finite. The bare
couplings and physical predictions are independent of the renormalization 
group scale $\mu$. Keeping the bare couplings fixed, we vary the RG scale
$\mu$ and use Eq.~\ref{eq:barerenorm} to determine the RG flow of the renormalized 
couplings $m^2(\mu),\lambda(\mu)$ and $y(\mu)$.

In Fig.~\ref{fig:RG} we plot the RG flow of $\lambda(\mu)$ and $y(\mu)$
using a lattice regulator for the integrals and with the arbitrary
choice $\lambda_0=0.1, y_0=0.3$. The RG scale $\mu$ is in lattice-spacing
units. When the RG scale is much lower than the cut-off ($\mu \ll 1$),
the renormalized couplings flow exactly as predicted by the 
continuum large-$N_{\rm F}$ RG equations
\begin{eqnarray}
y^2(\mu) &=& y_1^2/[1 - (N_{\rm F} y_1^2/4 \pi^2) \ln \mu] \nonumber \\
\lambda(\mu) &=& 12 y^2(\mu) + c_1 y^4(\mu).
\label{eq:continuumRG}
\end{eqnarray}
In the limit where the cut-off is infinite, $\mu \rightarrow 0$ and
the renormalized couplings $y(\mu)$ and $\lambda(\mu)$ vanish. As we
said, the theory is trivial, with no interaction when the regulator
is removed. In this regime ($\mu \ll 1$), continuum renormalized perturbation 
theory is perfectly valid, as the effects of the finite cut-off are negligible. 
When the RG scale is close to the cut-off ($\mu \sim 1$), the true RG flow of 
the couplings deviates from the continuum prediction. As $\mu$ increases, 
continuum perturbation theory predicts that $\lambda(\mu)$ becomes negative 
--- this is the vacuum instability. However for $\mu \gg 1$, Eqs.~\ref{eq:counter} 
and \ref{eq:zphi} show that  $\delta \lambda \rightarrow 0$ and $Z_{\phi} 
\rightarrow 1$. The renormalized couplings actually flow to the bare 
values $\lambda(\mu) \rightarrow \lambda_0, y(\mu) \rightarrow y_0$, exactly 
as shown in Fig.~\ref{fig:RG}. The continuum prediction that $\lambda(\mu)$ 
turns negative and the potential becomes unstable is incorrect, because the
cut-off dependence of the true RG flow has been neglected.

\section{TRUE HIGGS LOWER BOUND}

\begin{figure}[thb]
\vspace{12pt}
\epsfig{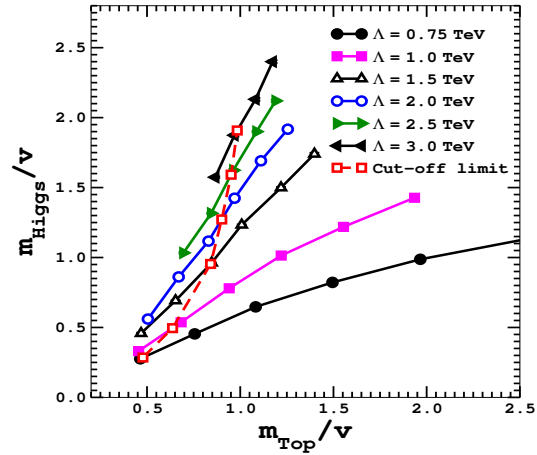}
\caption{Summary of lattice simulation results.}
\label{fig:simulations}
\end{figure}

The widely-accepted $m_{\rm Higgs}$ lower bound of Fig.~\ref{fig:PDG} is 
based on a vacuum instability appearing if the Higgs is too light. Using 
non-perturbative lattice calculations, perturbation theory, and the
large-$N_{\rm F}$ limit, we see that there is no vacuum instability.
The theory is trivial and the cut-off must be kept finite for the 
renormalized couplings not to vanish. The effective potential only 
appears unstable when the cut-off dependence of the renormalization 
procedure is ignored.

If there is no vacuum instability, is there an $m_{\rm Higgs}$ lower 
bound? As the theory is trivial, we require a finite cut-off to have a
non-trivial interaction. In the Higgs phase of the theory very close to the 
critical surface (as shown in Fig.~\ref{fig:phase}), the cut-off is
finite but large. For a fixed cut-off, upper and lower bounds for 
$m_{\rm Higgs}$ can be determined by exploring all allowed bare 
couplings \cite{Dashen:1983ts}. Moving away from the critical surface, the 
cut-off decreases and at some point the theory is completely dominated by 
cut-off effects and ceases to be physically acceptable .

\begin{figure}[thb]
\vspace{9pt}
\epsfig{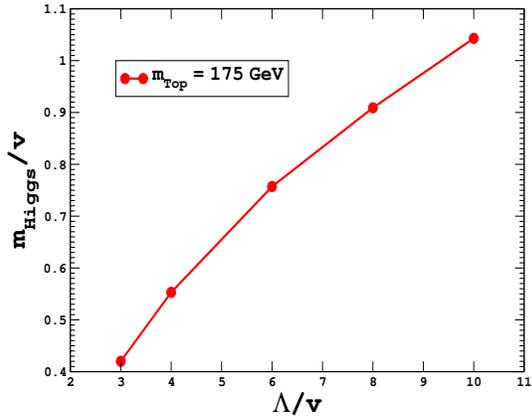}
\caption{Lower bound for $m_{\rm Top}=175$~GeV.}
\label{fig:lowerbound}
\end{figure}

We have used non-perturbative lattice simulations to explore the phase
diagram of the Higgs-Yukawa model with a single real scalar field
coupled to 2 flavors of staggered fermions ($N_{\rm F}
= 8$ continuum flavors). In Fig.~\ref{fig:simulations} we display a
summary of the results. All quantities are calculated in units of the
lattice spacing $a$. For illustrative purposes, the cut-off $\Lambda=\pi/a$ 
is converted into physical units using the Standard Model value 
$v = 246$~GeV. For example, $va=0.2$ corresponds to 
$\Lambda/v=\pi/0.2$~i.e.~$\Lambda \sim 4$~TeV. For fixed $m_{\rm Top}$
and $\Lambda$, we find that the smallest Higgs mass is generated when
the bare Higgs coupling $\lambda_0 \rightarrow 0$. The bare Higgs coupling
cannot be negative, otherwise the theory is not defined, as all 
functional integrals diverge. The solid curves in Fig.~\ref{fig:simulations}
correspond to $\lambda_0 = 0$ and $\Lambda$ fixed in physical units
(with three bare parameters and two constraints, this leaves one degree of 
freedom). With a finite cut-off, one must keep track of the finite cut-off 
effects. We use the ad hoc definition $m_{\rm ph}a=0.5$, applied to both 
Higgs and Top masses, as the smallest allowed correlation length, corresponding 
to the dashed line in Fig.~\ref{fig:simulations}. To the left of the dashed 
line, we expect the cut-off effects to be reasonably small and the theory 
to be physically acceptable. A similar constraint was used by L\"{u}scher and 
Weisz for the $O(4)$ Higgs sector, where the violation of rotational symmetry in 
Goldstone-boson scattering was investigated \cite{Luscher:1988uq}.

Interpolating the results sketched in Fig.~\ref{fig:simulations} we can extract 
the Higgs mass lower bound as a function of the cut-off for fixed physical Top 
mass. The curve in Fig.~\ref{fig:lowerbound} corresponds to the Higgs lower bound at
$m_{\rm Top}=175$~GeV. It is quite natural that the Higgs lower bound is attained 
when the bare Higgs coupling $\lambda_0 = 0$. Previous studies
of the pure Higgs $\lambda \phi^4$ theory showed that the $m_{\rm Higgs}$
upper bound is reached when $\lambda_0 = \infty$ \cite{Kuti:1987nr}.

\begin{figure}[thb]
\vspace{9pt}
\epsfig{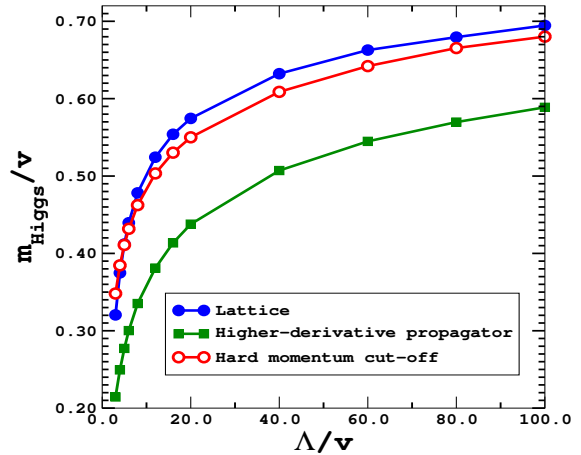}
\caption{Regulator-dependence of lower bound.}
\label{fig:regulator}
\end{figure}

One important consideration is that the lower bound is regulator-dependent.
Consider the $N_{\rm F} \rightarrow \infty$ limit of the Higgs-Yukawa model,
where perturbation theory becomes exact. We compare the Higgs lower bound,
calculated using three different regulators: a hard cut-off in the 
momentum integration, a lattice regulator, and a higher-derivative
propagator $k^2 \rightarrow k^2(1+k^2/\Lambda^2)^2$. In Fig.~\ref{fig:regulator} we 
plot the Higgs lower bound for a fixed Top mass. Even when the cut-off is quite
large, the lower bound varies by as much as 20\% among these three regulators. 
This is a feature of trivial theories which cannot be ignored --- when the 
cut-off is finite, not all quantities are universal. For any given regulator, 
one can calculate the Higgs lower bound to whatever desired accuracy. However, 
one cannot make arbitrarily accurate predictions which are regulator-independent.
There is an inherent ambiguity in the bound and one can at best estimate
the energy scale where the theory is no longer physically acceptable.

\section{PROPOSAL}

We have examined a toy Higgs-Yukawa model of one real scalar field coupled
to $N_{\rm F}$ degenerate fermions. Earlier studies of Higgs-Yukawa models
were reported in \cite{Lin:1993hp}. To make a quantitative statement 
relevant for the Standard Model, one needs to study a more physical model.
A realistic approximation would be an $SU(2)$ Higgs doublet (corresponding
to an $O(4)$-symmetric scalar field) coupled to a single fermion flavor
(the Top quark). Gluons should also be included, as the QCD coupling makes 
a significant contribution to the RG flow of the Top Yukawa coupling.
The remaining degrees of freedom of the Standard Model are expected
to play a negligible role in the lower bound. 
If it is only possible to calculate the lower bound via lattice
simulations, a gluon-Higgs-Top model is a very challenging system to 
explore. The only completely satisfactory way to represent a single
massless fermion flavor on the lattice is to use a chiral Ginsparg-Wilson 
fermion \cite{Ginsparg:1981bj},
which is computationally very demanding. Coupled to fluctuating scalar
fields, the positivity of the fermion determinant in the functional 
integral is not guaranteed, and must be examined in the Higgs phase
of the theory. If the probability distribution in the functional integral
can be negative, this could make numerical computations very
difficult (the so-call Sign problem). This is analogous to the situation in 
lattice QCD and how light a quark mass is possible in practical numerical simulations, 
which is very sensitive to the nature of the lattice fermion. Despite these 
challenges, we believe a realistic lattice study of a gluon-Higgs-Top 
system can make a significant and timely
contribution to current tests of the Standard Model.

\end{document}